\documentclass[preprint]{aastex}
\newcommand{\bc}{BRITE-Constellation}

\usepackage{graphicx}
\usepackage{natbib}

\shorttitle{\bc\ photometric nano-satellites}
\shortauthors{Weiss et al.}

\begin{document}

\title{BRITE-Constellation: Nanosatellites \\ for Precision Photometry of Bright Stars}

\author
{W. W. Weiss$^{1\dagger\star}$,
S. M. Rucinski$^{2\dagger}$,
A. F. J. Moffat$^{3\dagger}$, 
A. Schwarzenberg-Czerny$^{4\dagger}$, \\
{O. F. Koudelka$^{5\dagger}$},
C. C. Grant$^6$, 
R. E. Zee$^6$,
R. Kuschnig$^{1\dagger}$,
{St. Mochnacki$^{2\dagger}$},
J. M. Matthews$^{7\dagger}$, 
P. Orleanski $^{8\dagger}$,
A. Pamyatnykh$^{4\dagger}$,
{A. Pigulski$^{9\dagger}$}, 
J. Alves$^{1\dagger}$,
{M. Guedel$^{1\dagger}$}, 
G. Handler$^{4\dagger}$, 
\makebox{G. A. Wade$^{10\dagger}$,}
K. Zwintz$^{11}$, 
\and the CCD \& Photometry Tiger Teams$^{\ddagger}$
}

\affil{$^1$University of Vienna, Department of Astrophysics, Tuerkenschanzstr. 17, 1180 Vienna, Austria; 
$^2$Dept. of Astronomy and Astrophysics, University of Toronto, Canada;
$^3$Dept. de physique, Universit\'e de Montr\'eal;
$^4$Copernicus Astronomical Center, Warsaw, Poland;
$^5$Graz University of Technology. Graz, Austria;
$^6$Space Flight Laboratory, University of Toronto, Canada;
$^7$Dept. of Physics and Astronomy, University of British Columbia, Canada;
$^8$Space Research Center of the Polish Academy of Sciences, Warsaw, Poland;
$^9$Astronomical Institute, University of Wroc{\l}aw, Poland;
$^{10}$Dept. of Physics, Royal Military College of Canada, Ontario, Canada; \\
$^{11}$Instituut voor Sterrenkunde, KU Leuven, Belgium
}
\altaffiltext{$\dagger$}{Member of the \bc\ Executive Science Team (BEST)}
\altaffiltext{$\ddagger$}{M. Chaumont, S. Choi, Cordell Grant, T. Kallinger, J. Lifshits, B. Pablo, A. Popowicz, T. Ramiaramanantsoa, M. Rataj, P. Romano, J. Rowe, M. Unterberger, R. Wawrzaszek, G. Whittaker, T. Zawistowski, E. Zoc{\l}o\'nska  \& BEST}
\altaffiltext{$\star$}{email: werner.weiss@univie.ac.at}

\begin{abstract}
\bc\ (where BRITE stands for BRIght Target Explorer) is an international nanosatellite mission to monitor photometrically, in two colours, the brightness and temperature variations of stars generally brighter than mag(V)$\approx 4$, with precision and time coverage not possible from the ground. 
 
The current mission design consists of six nanosats (hence ÔÔConstellationÒ): two from Austria, two from Canada, and two from Poland. Each 7\,kg nanosat carries an optical telescope of aperture 3\,cm feeding an uncooled CCD.  One instrument in each pair is equipped with a blue filter; the other with a red filter.  Each BRITE instrument has a wide field of view ($\approx 24$ degrees), so up to about 15 bright stars can be observed simultaneously, sampled in 32\,pixel$\times$32\,pixel sub-rasters.  Photometry of additional fainter targets, with reduced precision but thorough time sampling, will be possible through onboard data processing.

The BRITE sample is dominated by the most intrinsically luminous stars: massive stars seen at all evolutionary stages, and evolved medium-mass stars at the very end of their nuclear burning phases.  The  goals of \bc\ are to (1) measure p- and g-mode pulsations to probe the interiors and ages of stars through asteroseismology; (2) look for varying spots on the starsÔ surfaces carried across the stellar disks by rotation, which are the sources of co-rotating interaction regions in the winds of the most luminous stars, probably arising from magnetic subsurface convection; and (3) search for planetary transits.
\end{abstract}


\section{History} \label{s:intro}                         
The birth of \bc\ can be traced directly to two developments: the Canadian microsatellite project MOST \citep{most1, matth2} and the beginning of the nanosatellite program at the Space Flight Laboratory (SFL) of the University of Toronto in the early 2000Ôs. The MOST mission (PI: Jaymie Matthews), proposed in 1997 and launched on 30 June 2003, is still delivering photometric data of the highest quality. Several members of the \bc\ Science Team are members of the MOST Science Team, and SFL is a MOST prime subcontractor, resulting in continuity of ideas and expertise -- both scientific and technical Ð from MOST to \bc. Soon after the inception of the nanosatellite (mass $<10$\,kg) program at SFL, one member of the MOST Science Team, Slavek Rucinski, put forward the idea of such a nanosat to study the brightest stars in the nightsky.  It was envisioned as the third in the series of SFL nanosats (https://www.utias-sfl.net/nano-satellites/) and most sophisticated, three-axis stabilized satellite in the series.

A workshop on nanosatellites was held in Vienna, Austria, in September 2004 to explore the technical and scientific potentials of this emerging technology, coinciding with an Austrian government initiative to improve the infrastructure of Austrian universities. A proposal for University of Vienna funding of the development of a space astronomy nanosatellite at SFL was submitted by the then Acting Chairman of the UniversityÕs Institute for Astronomy, Michel Breger. The proposal was approved in December 2005 and the satellite project was dubbed UniBRITE, with PI Werner Weiss. 

Independently, the Austrian Research Promotion Agency (FFG) announced a space technology program, to which the Graz University of Technology responded with a proposal to build in Austria (with technical support from SFL) another satellite as a twin to UniBRITE. This proposal was accepted in February 2006 and was named BRITE-Austria (TUGSAT-1), with PI Otto Koudelka. 

\begin{table}[h]
\caption{Launch and orbit information for the BRITE nanosats, where the entries for the Austrian and first Polish payloads are based on the actual launch, and for the Canadian and second Polish payloads are the planned parameters.  The designation is the official satellite name according to the United Nations Register of objects launched into outer space. F -- filter; T0 -- launch date; Orbit -- km above ground; Drift -- rate of drift from the initial ascending node in units of minutes per year, with ``dd" referring to a dusk-dawn orbit. \newline}
\begin{tabular}{l|l|c|l|c|c|c|c}
\hline
Designation & Name & F&Launcher & T0 & Orbit & descending & Drift\\
                   &           &        &            &      &  km        & node          &   \\
\hline
\multicolumn{2}{c|}{Owner: Austria} & & & & & & \\
BRITE-A & TUGSAT-1 & $\tilde{B}$ &PSLV-20 & 25 Feb. 2013 & $781\times 766$ & 18:00 & dd\\
BRITE-U & UniBRITE         & $\tilde{R}$ &PSLV-20 & 25 Feb. 2013 & $781\times 766$ & 18:00 & dd\\  
 \multicolumn{2}{c|}{Owner: Poland} & & & & & & \\
BRITE-P1  & Lem      & $\tilde{B}$    &DNEPR  &21 Nov. 2013 & $600\times 900$ & 10:30 & 100 \\
BRITE-P2  & Heweliusz& $\tilde{R}$   &China LM-4& Q2/2014 &    SSO/630\,km   &           &        \\
\multicolumn{2}{c|}{Owner: Canada} & & & & & & \\
BRITE-C2 & Montr\'eal&  $\tilde{B}$   & DNEPR  & Q2/2014& $629\times 577$ & 10:30 & 40 \\          
BRITE-C1 & Toronto&  $\tilde{R}$     &DNEPR  & Q2/2014& $629\times 577$ & 10:30 & 40 \\
\hline
\end{tabular}

\label{t:launch}
\end{table}
Adding additional nanosats added new capabilities to the BRITE mission.  For example, one satellite could observe a field when it is inaccessible to the other, or the number of fields could be doubled, or each satellite could be equipped with a different filter so two-colour observations become possible, or two satellites with the same filter could expand the time coverage.  The two-colour option was soon adopted, giving \bc\ a unique capability compared to the other space photometry missions MOST \citep{matth1}, CoRoT \citep{corot2}, and Kepler \citep{kepler1}, of which only MOST remains fully operational. 

Although the Canadian BRITE team made significant contributions to the scientific and technical development of the mission, attempts at securing funding for the Canadian BRITE nanosats were unsuccessful for several years. The first Canadian proposal in May 2005 was led by Slavek Rucinski; the second, in 2007, by Tony Moffat. Canadian Space Agency (CSA) funding for two Canadian BRITE nanosats was finally obtained in 2011. In the meantime, Poland had joined the project, thanks to an initiative led by Alexander Schwarzenberg-Czerny, resulting in funding in 2009 by the Polish Ministry of Science and Higher Education for another pair of BRITE nanosats. 

This led to the current configuration of \bc: three pairs of nanosats operated by three countries, observing in blue and red spectral bands.  Note, however, that the BRITE mission and its associated BRITE Executive Science Team (BEST) are open to other countries or organizations joining the Constellation with additional nanosats to expand the capabilities and/or the lifetime of the BRITE mission.

Both Austrian BRITE nanosats were launched together as secondary payloads on 25 February 2013, aboard a Polar Satellite Launch Vehicle (PSLV) C20 of the Indian Space Research Organization (ISRO) from the Satish Dhawan Space Centre in Sriharikota, India. UniBRITE and BRITE-Austria were inserted into dusk-dawn Sun-synchronous orbits with an orbital period of 100.3 minutes. The first Polish BRITE nanosat, named ``Lem" in honour of the famous Polish science fiction author Stanis{\l}aw Lem, was launched by ISC Kosmotras from Yasny, Russia, on 21 November 2013, aboard a DNEPR launch vehicle.  It was placed in a polar orbit of inclination $97.8^{\circ}$. Information on these launches and the resulting  orbital parameters, and on the expected parameters of the launches and orbits of the other Polish BRITE nanosat and the Canadian pair of nanosats is summarized in Table\,\ref{t:launch}.

\section{Science objectives and BRITE sample}      \label{s:object}     
\begin{figure}[h] 
\includegraphics[width=\textwidth]{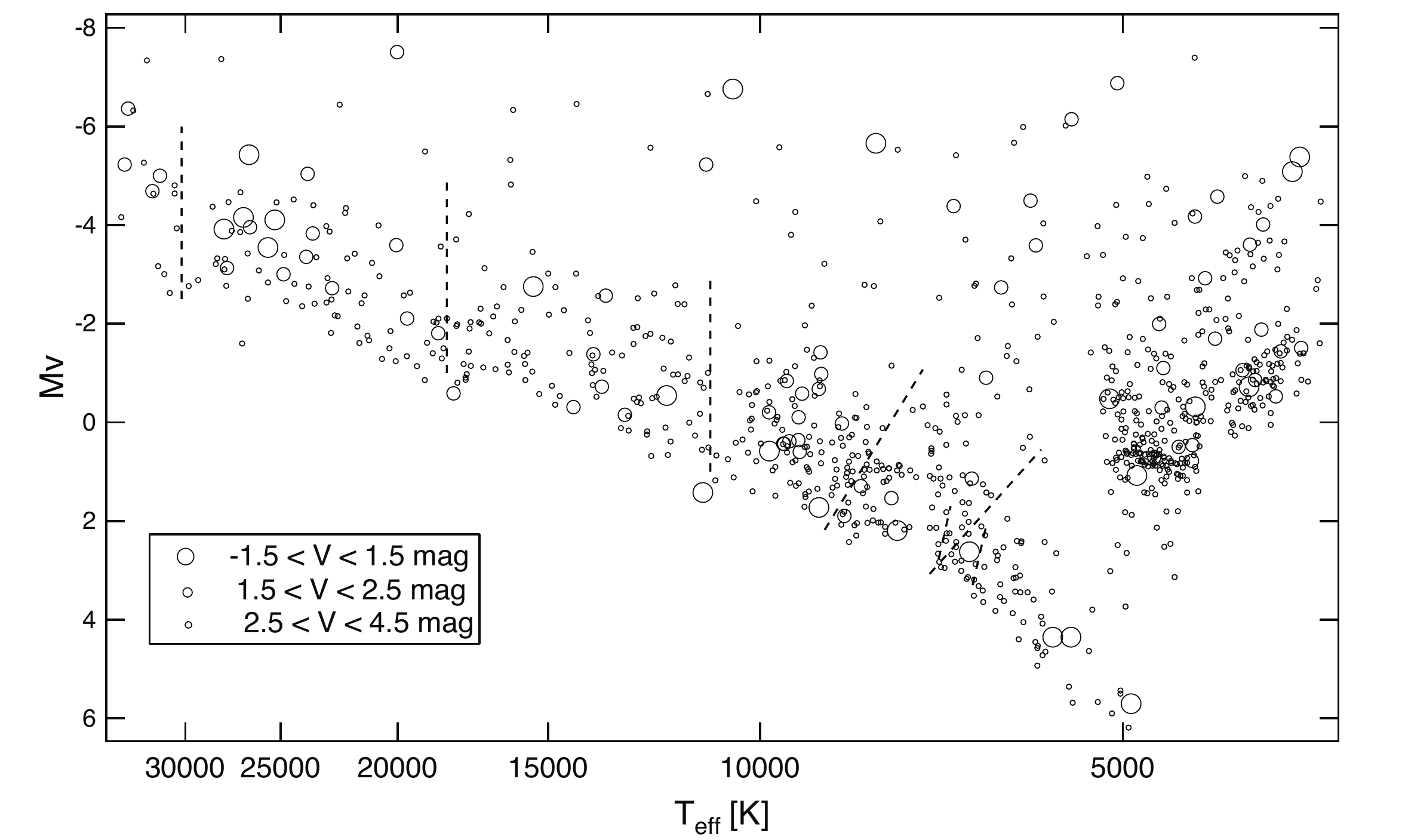}
\caption{Hertzsprung-Russell Diagram for stars brighter than mag(V)\,=\,4.5\,mag. This is the sample from which the primary \bc\ targets will be drawn. See text for more details.} 
\label{fig-hrd}
\end{figure}

The BRITE nanosats will survey a large part of sky, measuring brightness and temperature variations of the brightest stars on timescales ranging from a few minutes to several months (and perhaps years) via dual-broadband photometry resulting in a continuous precise photometric time-series unobtainable from the ground even with larger telescopes or existing telescope networks. Complementary spectroscopy of high spectral and time resolution can be obtained for these bright stars even with moderate-sized ground-based telescopes, which certainly is an advantage for getting telescope time. Such spectroscopic data are imperative for a full exploitation of the scientific information content of the BRITE photometry.

Why focus on stars of bright apparent magnitude?  Many of them are {\it intrinsically} very luminous. Hence, on average, the sample should include stars with initial masses greater than $\approx$\,8\,M$_\odot$, or medium-mass stars at the end of their lives when they become red giants and brighten by several magnitudes. There is a practical consideration too. Light curves of the brightest stars in the night sky obtained with a groundbased telescope often have poor precision because there are no suitably bright comparison stars at small angular separations available for differential photometry. The residual, incompletely corrected atmospheric extinction effects as a function of air mass introduce trends and additional scatter.

A Hertzsprung-Russell Diagram of the bright stars from which the \bc\ sample will be selected is shown in Fig.\,\ref{fig-hrd}. Groups of particular interest are, among others, OB (62), $\beta$\,Cep (29), CP (22), Be (20), EB (12), $\delta$\,Sct (7), HgMn (7), RR\,Lyr (3), roAp (1). However, the majority of stellar targets fall into two principal categories: 

(1) {\it Hot luminous H-burning stars} (O- to F-type) which represent almost half of the stars brighter than V\,=\,4.5\,mag). Time-resolved photometry of these stars has the potential to help solve three outstanding problems: (a) the sizes of convective cores in massive stars, (b) the influence of rapid rotation on their structure and evolution, and (c) the efficiencies of convection, mixing and overshooting. For the massive stars (O to early-B Main Sequence and all supergiants), \bc\ will investigate with asteroseisimic techniques their structure, which will cast light on the roles stellar winds play in supplementing the ISM \citep[e.g.,][]{moffat1, david1} and influencing future star and planet formation. 

(2) {\it Cool luminous stars} (Asymptotic Giant Branch (AGB) stars, cool giants and cool supergiants). Measurements of the time scales involved in surface granulation and differential rotation will constrain turbulent convection models. Oscillations in solar-type dwarfs have amplitudes of only a few parts per million (ppm) in luminosity \citep[e.g.][]{aerts1,houdek1}, which the precursor of \bc, MOST, detected for the first time in a star other than the Sun, in Procyon \citep{matth3, P1, P2}. p-mode pulsations in cool giants and g-modes in massive stars and cool giants, however, can have larger amplitudes measured in parts per thousand. 

\begin{figure}[h] 
\center
\includegraphics[width=0.6\textwidth, angle=0]{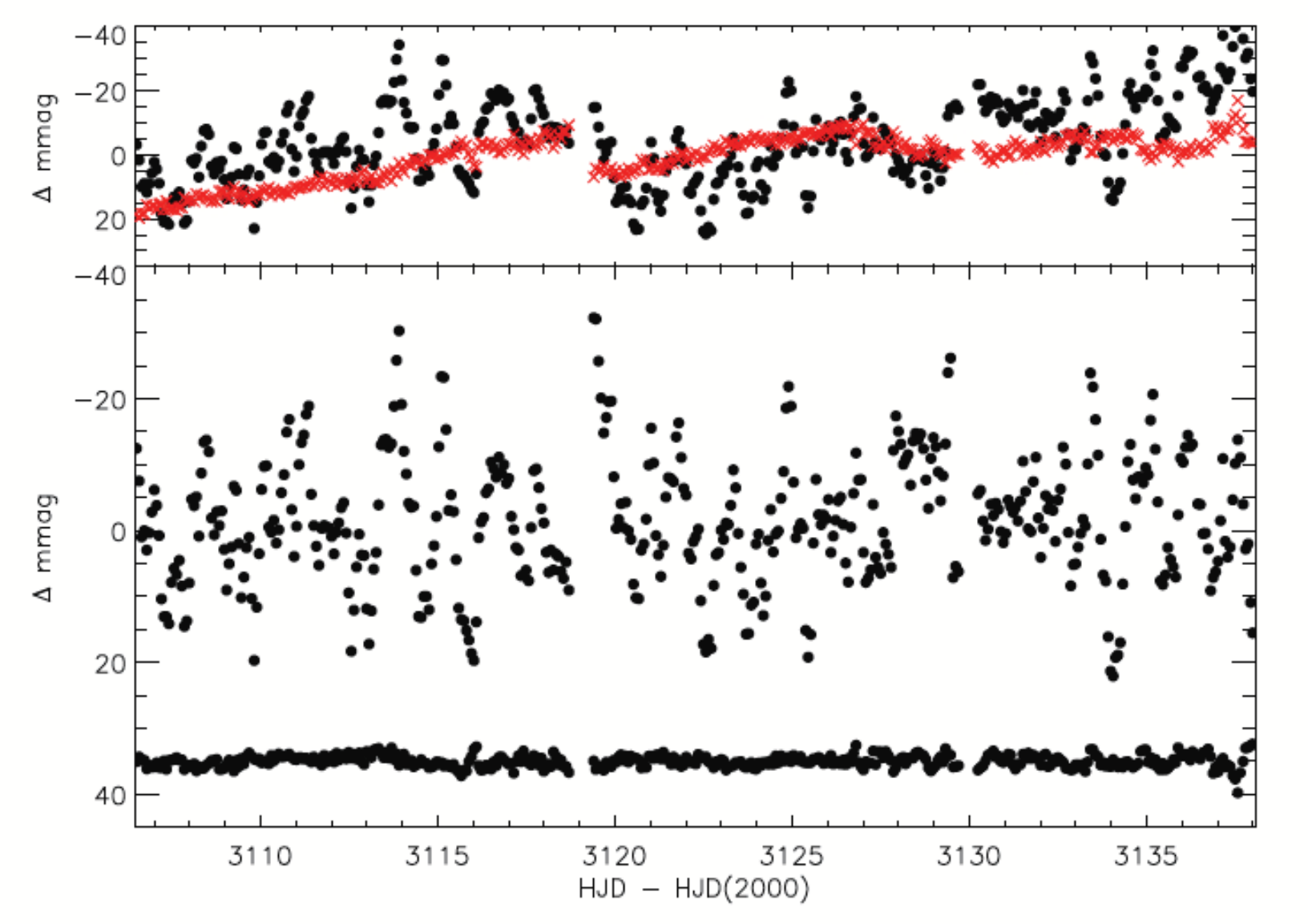} 
\caption{Orbital-mean MOST light curve of WR124 (WN8h) and a nearly constant comparison star in the same field observed simultaneously. Top panel: raw MOST light curves with comparison star shifted arbitrarily to superpose on that of WR124; bottom panel: light curves (now separated by an arbitrary amount for clarity) after removal of instrumental effects.  The zero-point in both cases refers to the mean of WR124.} 
\label{fig-WR}
\end{figure}

A few examples for space photometry contributing significantly to understanding of stars are: \\
\noindent
-- the runaway Wolf-Rayet star WR124\,=\,QR Sge of type WN8h, observed continuously for a month with MOST.  The light curve revealed for the first time a series of relatively cuspy peaks of varying intensity and lifetime (see Fig.\,\ref{fig-WR}, Moffat et al., in prep.).  This plot shows a dominating frequency at about 0.18\,cd$^{-1}$ (5.5-day period) with variable amplitude. \\
\begin{figure}[h] 
\center
\includegraphics[width=0.8\textwidth]{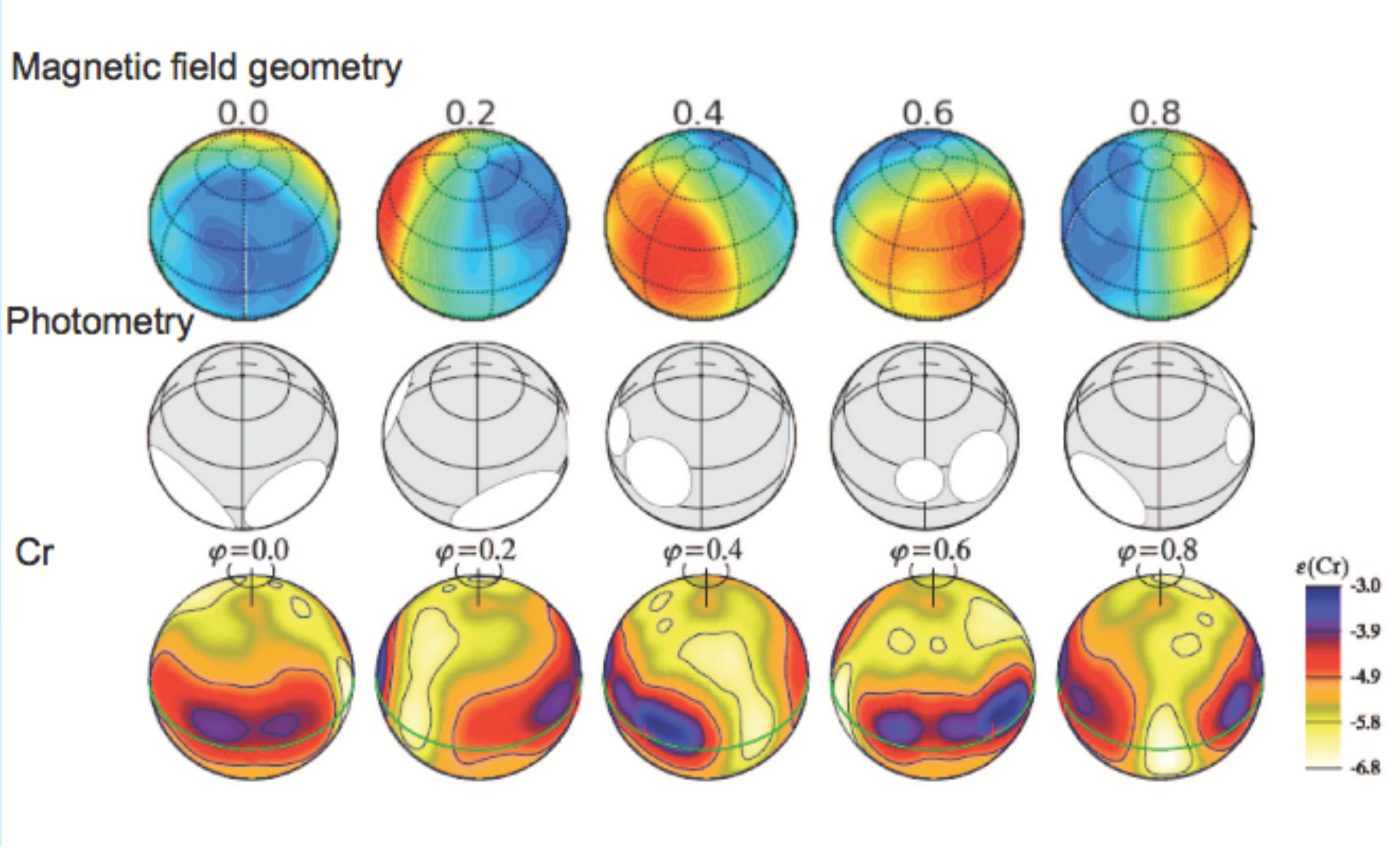}  
\vspace{-9mm}
\caption{Surface Doppler Imaging of the CP star HD\,50773. Top: surface magnetic field distribution derived with spectropolarimetry; Middle: photometric surface imaging with identification of bright spots determined from a CoRoT high precision, long time base light curve; Bottom: surface distribution of chromium derived from spectroscopy. From \cite{lueft1}.} 
\label{fig-doppler}
\end{figure}
-- classical pulsating stars (e.g. $\delta$\,Scuti stars) whose oscillations can be characterised down to low amplitudes \citep[e.g.,][]{zwintz1}, \\
-- stars in clusters and association which share a common age but have a range of masses  \citep[e.g.,][]{breger1}, \\
-- stars with spots in their photospheres \citep[e.g.,][]{lueft1, nesva1} and their temporal evolution (see Fig.\,\ref{fig-doppler}). \\
-- granulation signatures across the HR Diagram \citep[e.g.,][]{kall2}. \\
-- Targets of Opportunity, such as bright comets, novae and supernovae. \\
-- As important as the stellar variability will be for BRITE science,  non-variability may also teach us much. BRITE may enlarge the base of stars measured to be constant (within low noise limits) to serve as photometric standards for other studies. \\
-- Last, but not least, we see a chance to detect exoplanet transits for bright stars \citep[see][]{pass1, pass2}.

Complementary ground based, in particular spectroscopic observations are needed to fully exploit the scientific content of space photometry. It was therefore decided already in an early stage of the project to establish a Ground Based Observing Team (GBOT), with G. Handler, A. Pigulski and K. Zwintz (chair) as managers.

\section{Sky and time coverage}         

\begin{figure}[h] 
\includegraphics[width=\textwidth]{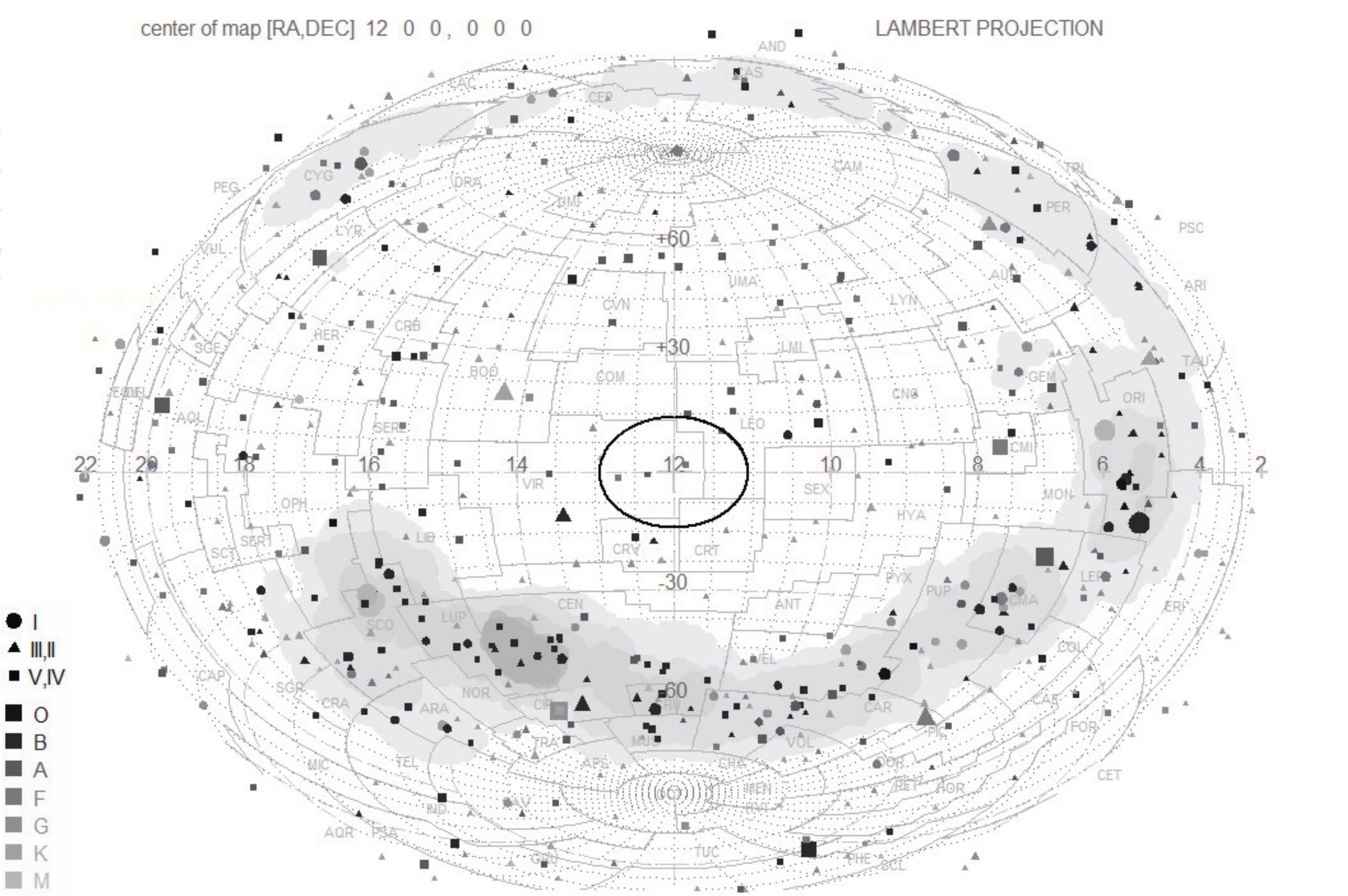}
\caption{A map of the sky including all stars brighter than V\,=\,4.5\,mag. Center circle: An example of a BRITE 24-degree field of view.} 
\label{fig-sky}
\end{figure}

The intrinsically luminous, apparently brightest stars are not distributed evenly across the sky, but are concentrated along the plane of the Milky Way. Figure\,\ref{fig-sky} shows a map of the sky including all stars brighter than V = 4.5\,mag. Spectral types are coded and the shapes of the symbols represent the luminosity classes. The grayscale band is neither surface brightness nor star density of the Milky Way, but target density for the 24-degree field of view (FOV) of one BRITE instrument. The darker the grayscale, the larger the number of targets -- up to 16 -- observable in a single pointing.

One factor that has contributed to the extraordinary successes of MOST, CoRoT and Kepler is the very clean Fourier spectral window of the observations with a long nearly continuous time baseline.  This results in unprecedented accuracy of the determined frequencies and improving the amplitude precision in Fourier space, thanks to a high duty cycle of typically 95\% for target stars observed during spans of months and up to years.

\bc\ has the potential to observe in essentially all directions of the sky, so targets will generally be occulted by Earth during part of any one satellite orbit which results in periodic gaps in data sets. The BRITE solution was to make BRITE a Constellation of several nanosats equipped with the same instrumentation in different orbits, so one BRITE nanosat can fill the data gap of another nanosat. 

The longer the continuous time series, the better the resolution of oscillation and rotational frequencies in the stellar data.  The goal of \bc\ is to obtain time coverage of up to years for some targets. The simplest estimate of the frequency resolution is the Rayleigh criterion (1/T, where T is the total length of the time series). A more realistic estimate of the frequency resolution incorporates the signal-to-noise ratio of the data \citep[see][]{ASC,kall3}.

\section{Two passbands}       \label{s:passband}    

One of the challenges in asteroseismology of stars which are not pulsating in high overtone modes like the Sun and cool giants, is mode identification.  Each pulsation mode can be described by a specific eigenfunction and the different eigenfunctions are sensitive to different depths within the star. This enables an asteroseismologist to `peel away' the layers of a star like an onion and determine its internal structure.
 
\begin{figure}[h] 
\center
\includegraphics[width=0.7\textwidth]{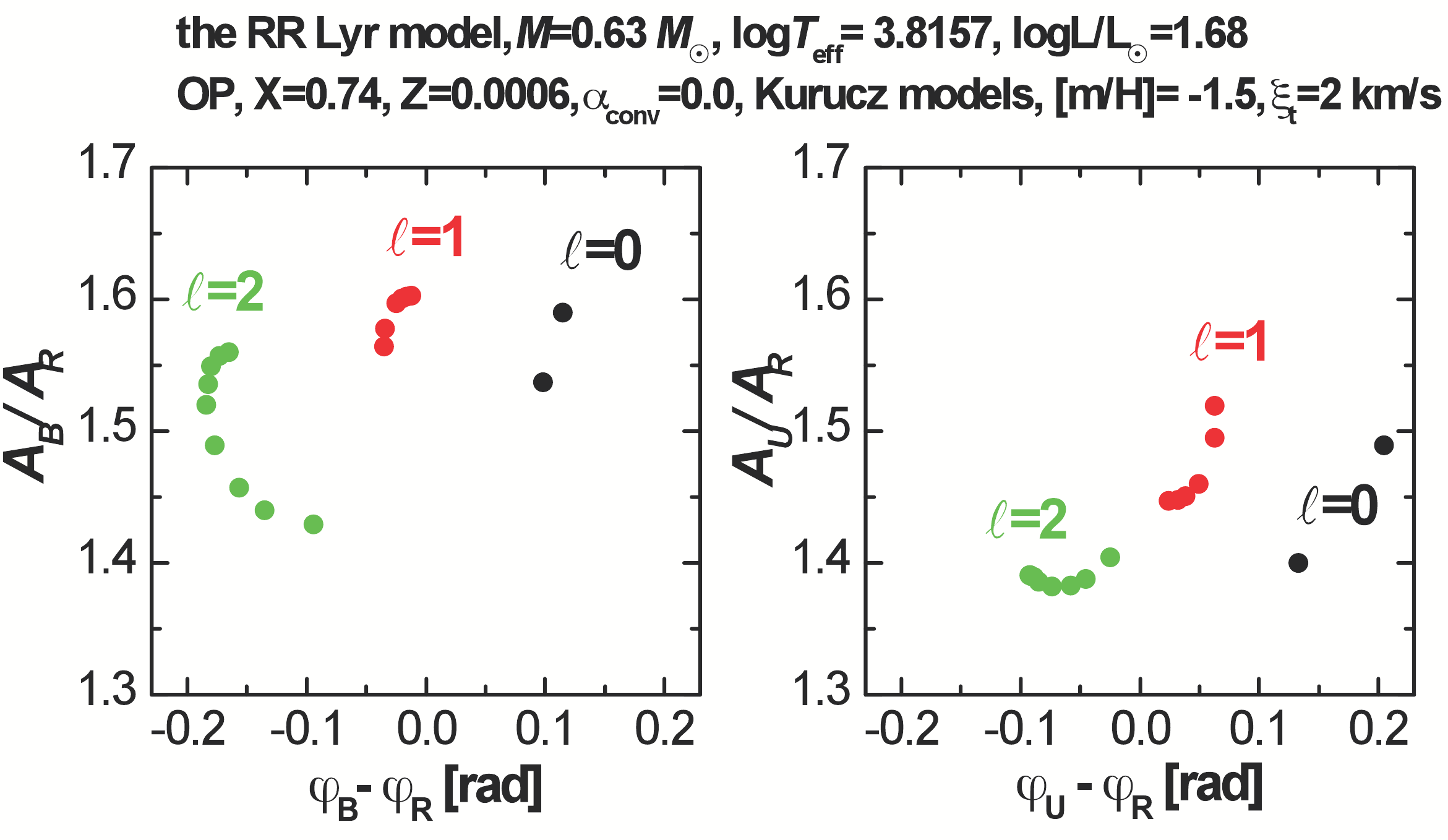}
\caption{Pulsation phase differences and amplitude ratios for radial and nonradial modes ($\ell = $ 0 to 2) of different overtones for a typical RR Lyrae pulsator, computed for the BRITE blue ($\tilde{B}$) and red ($\tilde{R}$) filters and a possible UV filter. Pulsation modes with increasing radial node numbers are plotted with the same colour. (Figure courtesy of Jadwiga Daszy\'nska-Daszkiewicz.)} 
\label{fig-rrlyr}
\end{figure}

Unfortunately, different modes may result in very similar frequencies and it is therefore important to attribute to each observed frequency the correct mode geometry. Solar and solar-like asteroseismology can take advantage of regular patterns seen in the oscillation frequencies in the high-overtone asymptotic regime \citep[e.g.,][]{kall1}. So-called avoided crossings can disrupt this regular pattern and render asymptotic fits based on large and small frequency spacings plotted in echelle diagrams \citep[e.g.,][]{lenz1} untrustworthy. Even moderate stellar rotation introduces asymmetries and coupling, complicating even more the patterns in the eigenfrequency spectrum \citep[e.g.,][]{pena1}. This is a particularly important effect in the BRITE sample, since most of our massive-star targets near the main sequence will have high rotation speeds. 

Measuring the amplitudes and phases of the oscillations in different ranges of wavelength offers one way to recognise independently the pulsation mode geometry through linearization of the light variations of a non-radially pulsating star \citep{jagoda1, jagoda2}, as illustrated in Fig.\,\ref{fig-rrlyr}. The ratio of amplitudes of different pulsation modes measured through different filters (in this case BRITE blue and red, and a possible UV filter) is plotted against the phase shifts of the pulsation modes measured in those passbands.  In this parameter space, the radial and two low-degree nonradial modes are clearly distinguished from each other.

\begin{figure}[h] 
\center
\includegraphics[width=\textwidth]{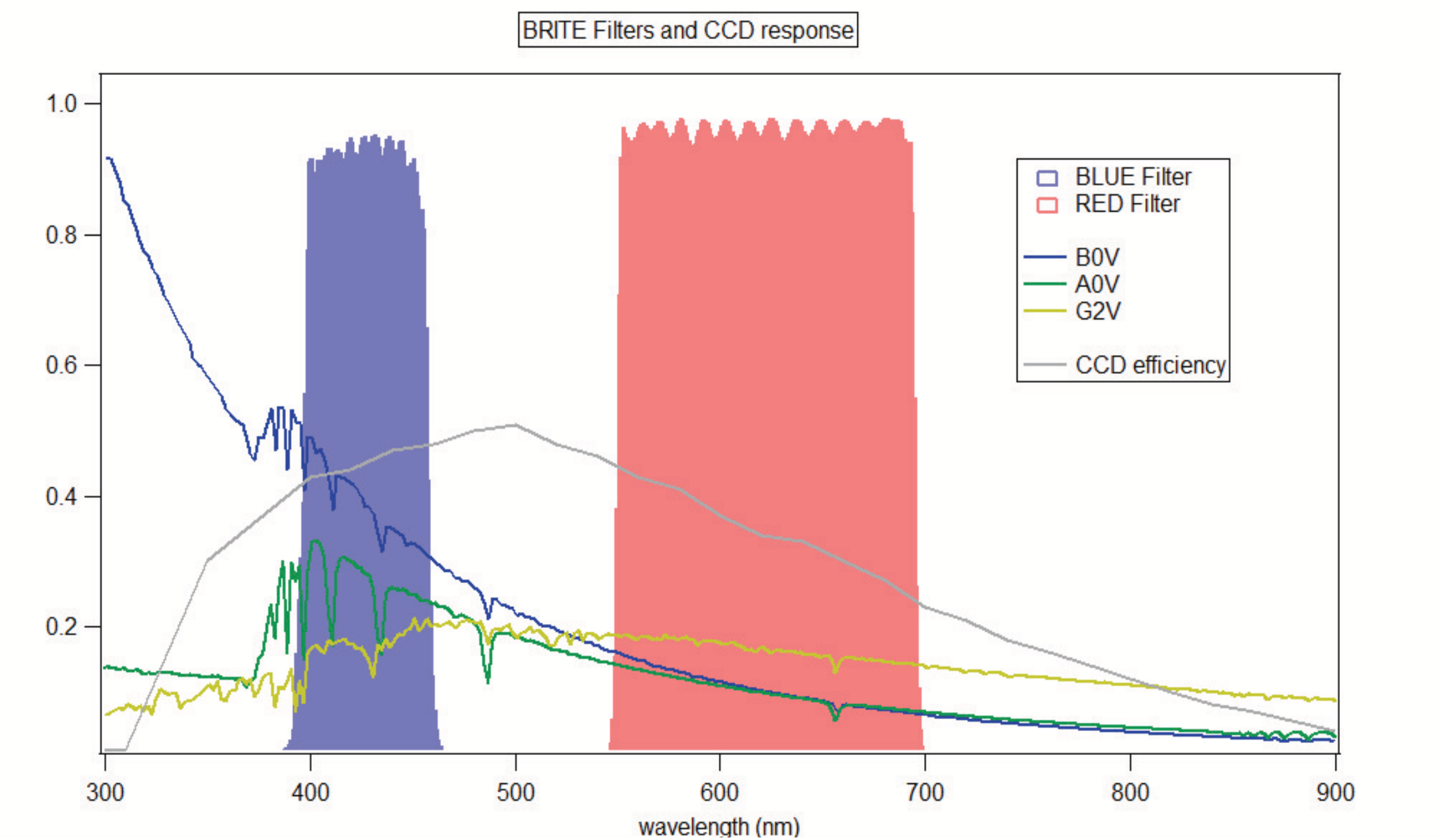}
\caption{Transmissions of the blue ($\tilde{B}$ for BRITE-Austria) and red ($\tilde{R}$ for UniBRITE) multilayer interference filter, compared to the relative spectral energy distributions of B-, A- and G-type stars, and the response of the BRITE CCD detector.} 
\label{fig-filter}
\end{figure}

The main practical considerations for the design of the BRITE two-colour system were: (1) matching the transmission of the BRITE instrument optics and the response of the BRITE CCD detector; (2) having broad passbands to yield good photon statistics despite the small aperture of each BRITE telescope; (3) giving the passbands the largest possible separation in wavelength, with minimum overlap; and (4) the desire to achieve comparable photon statistics for measurements of an ``average" star in the BRITE sample (near spectral type A0) through the two filters. These considerations led to the designs of the blue and red interference filters, whose transmissions are shown in Fig.\,\ref{fig-filter}, compared to the spectral energy distributions of an A0 dwarf, a B0 dwarf and the Sun (a G2 dwarf), as well as to the quantum efficiency of the BRITE detector.

\section{Photometry at the BRITE focal plane}         \label{s:photom}     

There are three different ways in which BRITE photometry is collected and initially processed:

(1) {\it Subraster photometry}: Up to 15 subrasters (each with dimensions of $32\times 32$ pixels, possibly reduced to $24\times 24$ pixels with the best telescope pointing performance) centred on target stars and sky background fields are routinely read out with a typical integration time of seconds. These subrasters can be co-added on board before being downlinked to Earth, resulting in an effective integration time of a few minutes up to 45\,minutes per BRITE orbit. 

(2) {\it Onboard processing} can be implemented in the future on request: Up to about 100 subrasters ($32\times 32$ pixels each) will be typically read out once every few seconds. The contents of each subraster will be represented by a set of 10 numbers calculated on board the satellite: the mean intensity values for pixels at the four corners of the subraster (to characterize the background), the sum of pixel values within two different concentric circles centred on the target in the subraster (aperture photometry), and the medians and standard deviations of these values. 

(3) {\it Full frame measurements}: At the beginning and end of an observing run of a given field, a complete CCD image consisting of 11 Mpixels (resulting in a file size of about 20\,MB) is transmitted to the ground.  The purpose of these full frames is to test and track cosmetic changes in the CCD (e.g., warm pixels resulting from radiation damage).  Downlinking just one such frame needs about 3 hours of telemetry, requiring up to 18 passes over a single ground station, which is equivalent to more than 3 days. Hence, full frame transfers, while valuable, will be relatively rare during the \bc\ mission.

The available numbers of subrasters (presently a few 10's per field) and on-board processed subrasters (presently up to 100 per field) depend mainly on the telemetry downlink capacity and the number of ground stations in the BRITE network. An efficient lossless compression algorithm developed by Franz Kerschbaum and Roland Ottensamer at the University of Vienna (private communication) is expected to at least double the number of subrasters that it will be possible to save and downlink to Earth.

\section{BRITE hardware and mission operation}       \label{s:hw}          

In the microsat philosophy first implemented by MOST, nano- and microsatellites are designed, built and tested according to the principle that reliability is increased when design complexity is decreased. In other words, simple satellites may be intrinsically more reliable, and as a bonus, less expensive. Moving parts are avoided wherever possible. This policy reduces considerably the cost, hence helping make a mission to perform pioneering space research accessible even to individual universities and research institutes.  Note that MOST employs mostly non-space-qualified electronics, based on a mission risk analysis, and it is still operating after more than 10 years, despite an required 1-year mission life.

Adopting this approach, it was decided to have no moving parts on each BRITE nanosat except for the Attitude Control System (ACS) reaction wheels.  With no filter wheel to change filters (or an expensive compact beam splitter), each BRITE carries a single fixed filter, and it is the constellation of satellites with different filters that makes the required time coverage in two passbands possible.

\begin{figure}[h] 
\center
\includegraphics[width=0.6\textwidth]{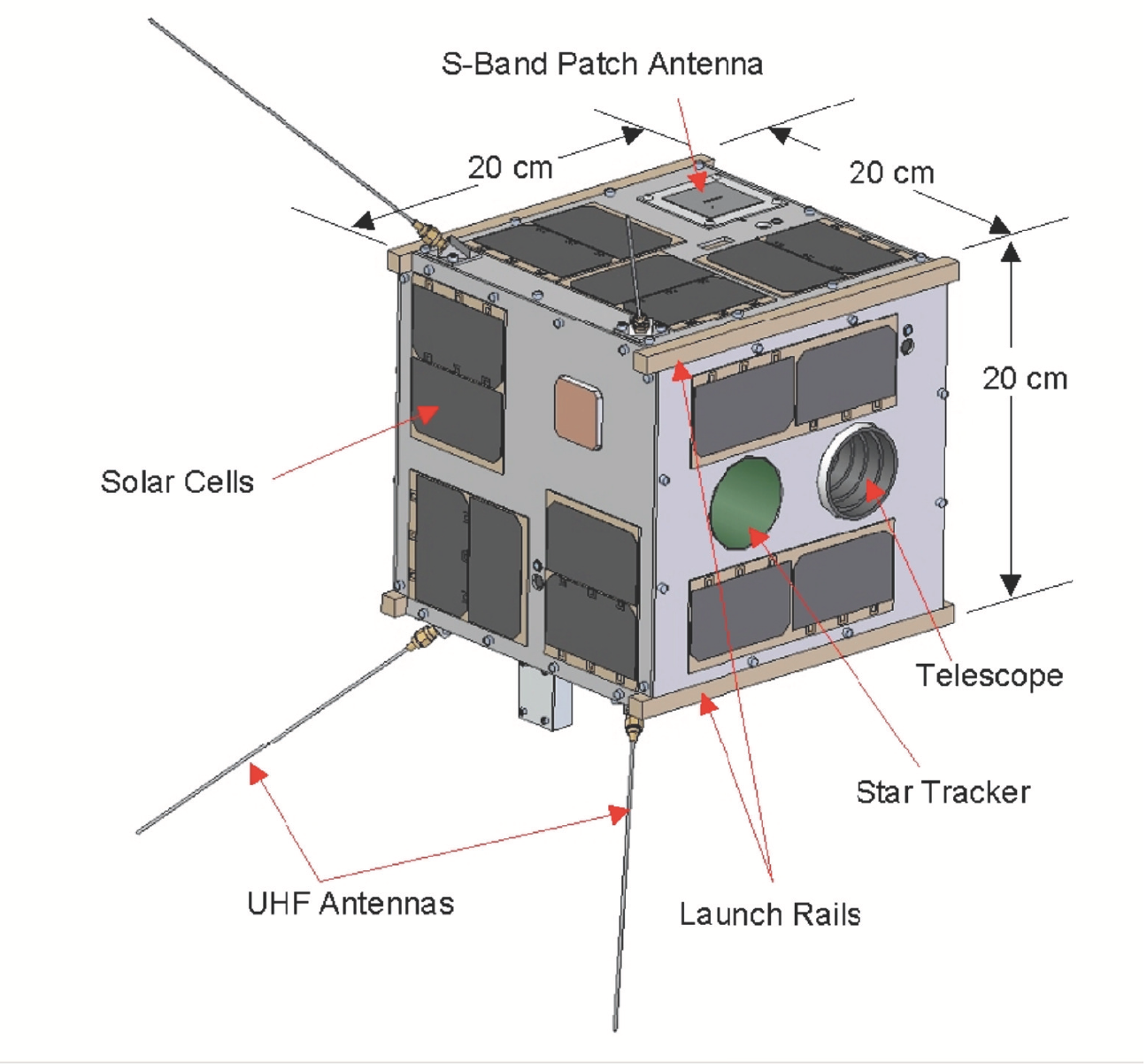}
\caption{Schematic of a BRITE nano-satellite. The launch rails are where the satellite bus interfaces with the UTIAS/SFL XPOD ejection system which separates the satellite from the launch vehicle.} 
\label{fig-bus}
\end{figure}

\subsection{BRITE bus}     \label{ss:bus}

The dimensions and principal exterior features of a BRITE nanosat are shown in Fig.\,\ref{fig-bus} and its top-level characteristics are listed in Table \ref{t:bus}.  The satellite is a cube 20 cm on a side, weighing 7 kg, with thin antennas and a magnetometer boom that extend from the bus. The main subsystems of the BRITE bus are:

(1) {\it Attitude determination and control}: The satellite Attitude Control System (ACS) is based on SFL high performance ACS technology and provides  {\sl rms} stability of approximately 1--1.5\,arcminutes (or 2--3 pixel at the CCD) according to the original mission scientific requirements. In fact, during on-orbit commissioning, better pointing performance (mean variation of 1.5 pixel) was achieved by both BRITE-Austria and UniBRITE. 

\begin{figure}[h] 
\center
\includegraphics[width=0.7\textwidth]{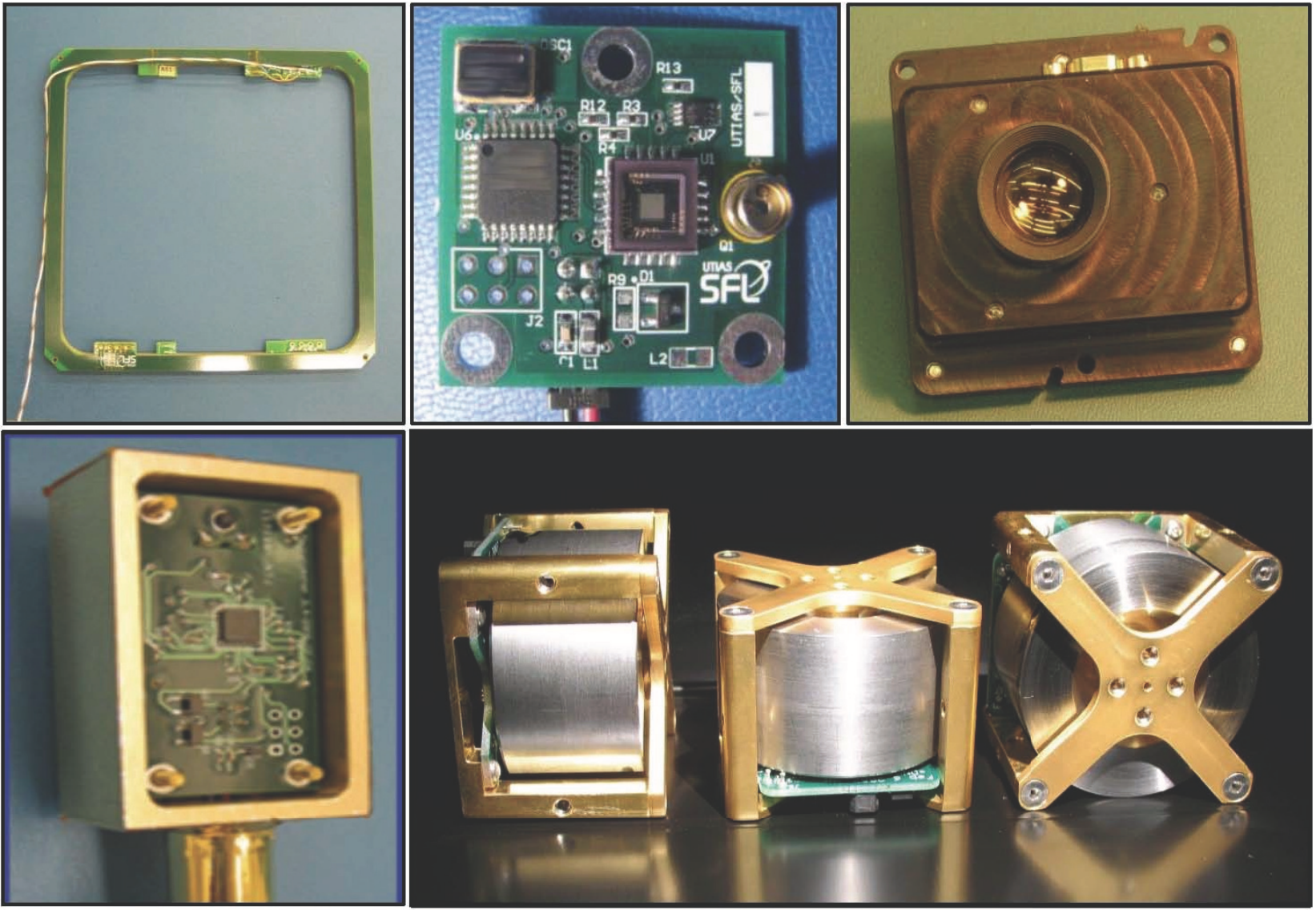}
\caption{Attitude control hardware. Top left: Magnetorquer coil, centre: Sun sensor, right: Startracker; Bottom left: Magnetometer, right: Reaction wheels} 
\label{fig-ACH}
\end{figure}

Each BRITE nanosat houses 3 orthogonal reaction wheels (developed by the company Sinclair Interplanetary in collaboration with SFL) and three orthogonal vacuum-core magnetorquer coils for 3-axis pointing and momentum dumping. Attitude determination is provided by a magnetometer and six SFL-developed Sun sensor packages, each of which is equipped with coarse and fine Sun-sensing elements. Figure\,\ref{fig-ACH} shows photos of the four main components of the ACS hardware. The bus also carries a nanosatellite star tracker developed by the company AeroAstro. (AeroAstro was the only supplier available when designing the Austrian pair of BRITE nanosats, and unfortunately AeroAstro is no longer in business.) The tracker is designed to point the satellite with an accuracy of  better than 1.5\, arcminute {\sl rms} (see also Section\,\ref{s:comm}).  The Canadian and Polish pair of nanosats are equipped with star trackers developed by Sinclair Interplanetary, Ryerson University (Toronto) and SFL, which are designed to give even better pointing performance than the AeroAstro design. 

(2) {\it Computers}: Each satellite houses a main onboard computer (OBC), and an attitude determination and control system (ADCS) computer. These computers have the same design, built around an ARM7 processor operating at $\approx40$\,MHz. The computers each contain 256\,KB of flash memory for code storage, 1\,MB of EDAC-protected SRAM for storage of instrument data and engineering telemetry, and 256\,MB of flash memory for long-term storage of telemetry and instrument data. The satellite also has a third OBC (of the same design as the main and ADCS OBCs) dedicated to operating the instrument (primarily its CCD camera). Communication between the instrument and the payload OBCs is asynchronous serial, with a maximum data transfer rate of 115.2\,kbps.

(3) {\it Power}: The power system provides an unregulated bus voltage of nominally 4.0\,V  and uses direct energy transfer from high-efficiency triple-junction solar cells to the satellite systems. A 4.8\,Ah rechargeable lithium-ion battery provides power for use during eclipses (when the Sun is below the Earth horizon as seen by the satellite) and periods of peak power usage. The nominal design of the BRITE bus allows for at least six solar cells on each face, for a maximum instantaneous power generation of $\approx10$\,W. The worst-case minimum power generation (when the Sun is visible to the satellite) is $\approx5.4$\,W.

(4) {\it Communications}: The command uplink to the satellite is handled on board by an SFL-developed UHF receiver. An S-band transmitter (also developed at SFL) provides the primary downlink for data and telemetry.  It can function at data rates of 32 -- 256\,kbit\,s$^{-1}$. Each radio system is housed in a separate enclosure to minimize noise and interference. The UHF transceiver communicates with the ground via a pre-deployed quad-canted monopole antenna array. The uplink data rate is 9.6\,kbit\,s$^{-1}$. The S-band transmitter uses two $5.5\times 5.5$\,cm SFL-developed patch antennas installed on opposite sides of the spacecraft providing near omnidirectional coverage.

(5) {\it Thermal Control}: Satellite and instrument thermal control is primarily passive. Thermal coatings and tapes are used on the external spacecraft surfaces to keep the average satellite temperature between 10$^{\rm o}$C and 40$^{\rm o}$C during an orbit. Partial thermal control of individual components is achieved using a variety of techniques including thermal isolation and heat sinking.

\begin{table}[h]
\caption{BRITE satellite bus specifications. \newline}
\begin{tabular}{l|l||ll|l}
\hline
Satellite Specification & Value & &Satellite Specification & Value \\
\hline
Volume & $20\times 20\times 20$\,cm & &Power  & 5.4\,W to 10\,W \\
Mass    & 7.0\,kg & &Bus Voltage &4.0\,V (nominal) \\
Attitude Determination & 10\,arcseconds & &Battery Capacity  & 5.3\,Ah \\
Attitude Control Accuracy & better than 1.0$\deg$ & &Data Downlink  & up to 256\,kbit/s  \\ 
Attitude Control Stability  &  1 arcminute  {\sl rms}         & &Data Uplink      & 9.6\,kbit/s \\       
Payload Data Storage&  up to 256\,MB  & &  &  \\
\hline
\end{tabular}
\label{t:bus}
\end{table}

\subsection{BRITE optics, image PSF and detector}     \label{ss:optic}     

\begin{figure}[h] 
\center
\includegraphics[width=0.7\textwidth]{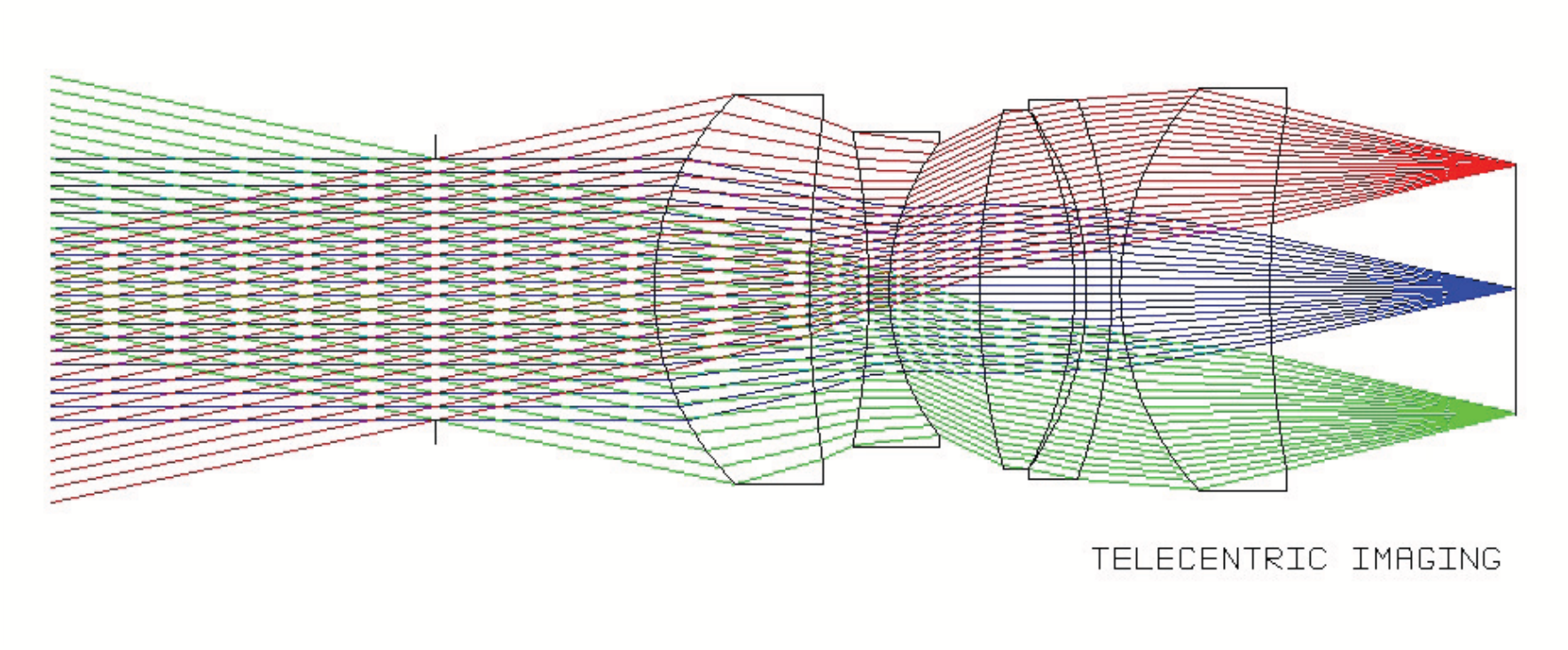}
\caption{BRITE optical layout and three sets of ray tracings (for an on-axis image and images $\pm12\deg$ off axis at the edges of the unvignetted field).} 
\label{fig-lens}
\end{figure}

The instrument consists of a small lens-based telescope and an uncooled CCD detector. The telescope objective (slightly different for the blue- and red-filter versions) has an aperture of 30\,mm and effective focal length of 70\,mm.  There are 5 lenses in the telecentric optical design (see Fig.\,\ref{fig-lens}), with a nearly unvignetted FOV of 24\,degrees. Only 4 lenses are in the second Polish BRITE, named ``Heweliusz".  The focal plane scale is about 27\,arcsec per pixel (where the pixels are 9 microns on a side). The optics for all the BRITE satellites, except for Heweliusz, were designed by Ceravolo Optical Systems (Canada), with emphases on (1) reducing stray light from scattered Earthshine, (2) minimizing vignetting, and (3) allowing easy physical access to the filter. The forward pupil design allows use of an efficient short external baffle so the distance from the end of the baffle to the CCD focal plane is only 17\,cm. The optics of Heweliusz were designed by the Space Research Centre (Poland), with the goal of achieving a smoother PSF profile.

The BRITE detector is a KAI-11002M CCD manufactured by Truesense Imaging Inc. (formerly Kodak Image Sensor Solutions), which is a full 35-mm format ($36\times 24$\,mm) detector with 11 million square pixels (each $9\times 9\,{\mu}$m).  The KAI-CCD was chosen for its large format, low dark current, relatively low readout noise, good quantum efficiency, and because it employs microlenses with interline transfer readout to replace a mechanical shutter. A CMOS detector was originally considered for BRITE, but tests showed that candidate CMOS detectors produced excessive fringes with the BRITE optics. 

The CCD is not actively cooled, but thermal simulations of the BRITE bus in typical orbits indicated that the average temperature near the CCD focal plane will be 10$^{\rm o}$C to 40$^{\rm o}$C.  Commissioning data from the first two BRITE nanosats have proven those simulations were accurate.  Sensors monitor the temperature of the CCD and its support electronics. There is a heater to provide additional stabilization of the detector temperature to within 0.1$^{\rm o}$C but, to conserve power, it will not be used unless it is shown to be needed. 

The goal of \bc\ is highly precise photometry and not the production of sharp stellar images. The photometric requirement calls for a point spread function (PSF) which is as smooth as possible, illuminating an area of up to about $8\times 8$ pixels on the CCD.  A larger PSF would reduce the noise due to satellite pointing wander and to flat-fielding errors, while also permitting higher S/N total flux levels (like in the Fabry mode of MOST). However, a larger PSF increases the risk of contamination of the target light by background stars due to blending. The choice of optimum de-focus of the optics was made based on numerical simulations, first with a pointing jitter model developed from experience with MOST's attitude control system, coupled with models of intra-pixel and pixel-to-pixel CCD sensitivity variations. The conclusions were later confirmed by pointing jitter models specific to the BRITE ACS design. 

\begin{figure}[h] 
\center
\includegraphics[width=0.4\textwidth]{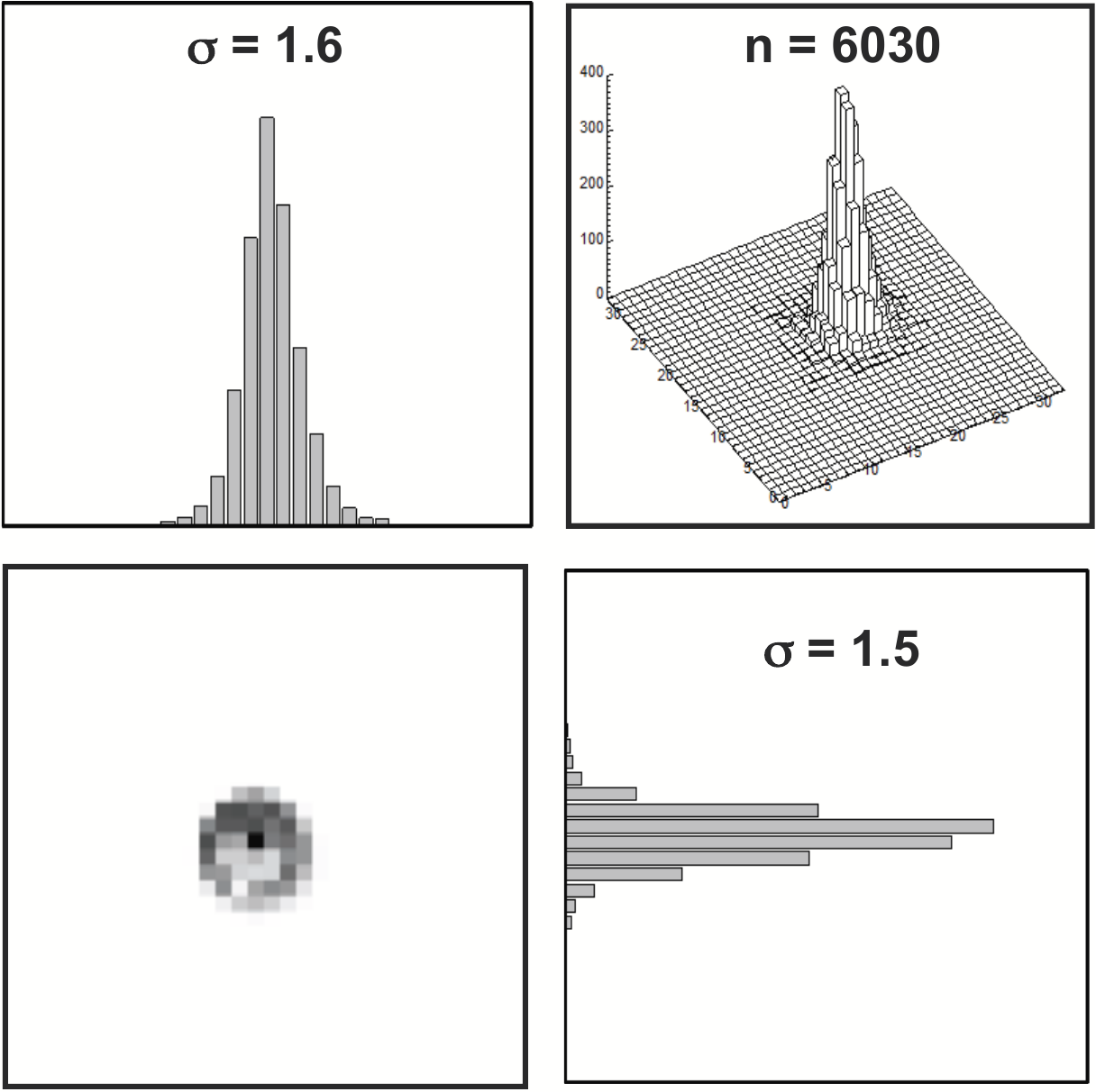}
\caption{BRITE-Austria point-spread-function of $\eta$\,Ori. Pointing accuracy of BRITE-Austria in X (upper left panel) and Y (bottom right panel) accumulated during 6030 individual subframes and after repeated acquisition of the field during 15 days; $\sigma$ in units of pixels, bin-size corresponds to a pixel.}
\label{fig-pointing}
\end{figure}

Figure\,\ref{fig-pointing} shows an actual PSF of the star $\eta$\,Ori observed with BRITE-Austria during commissioning. The two histograms illustrate the pointing accuracy of BRITE-Austria in X (upper left panel) and Y (bottom right panel) accumulated in a total of 6030 subframes.  Included are not only deviations during individual orbits, but also the accuracy to which the same field was acquired repeatedly during the 15 days of observation. In the case of the Polish red-filter BRITE instrument, ``Heweliusz",  it was found that the de-focused 4-lens system gives a comparable PSF spread, but one which is more Gaussian in shape. 

Since there are no moving parts (other than the spinning reaction wheels), there is no lens cover for the telescope objective that can be opened and closed.  Analyses and tests have demonstrated that, if the Sun were to cross the telescopic field of view of the tumbling nanosat, no damage to the CCD would result.  Ground tests of this scenario were performed with the engineering model (in vacuum behind an optical window) by directing sunlight with a mirror onto the instrument aperture.  The temperature at the CCD focal plane stayed well within the tolerances of the device.  Indeed, during the tumbling phase of BRITE-Austria after deployment in orbit, frames were taken with the Sun drifting through the FOV with no damage to the CCD.

Early observations during commissioning reveal that our choice of CCD, well suited to the mission in so many respects, had an unexpected cost: a higher-than-expected sensitivity of the CCD to radiation.  This has resulted in more bad (``warmÒ) pixels than expected. This is combined with the challenges of operating relatively warm CCDs of varying temperature, and highly non-symmetric PSFs especially near the outer edges of the field of view. We are currently in the process of exploring and mitigating these problems.

\subsection{Ground stations}     \label{ss:ground}    

The ground station and mission control centre for BRITE-Austria was established at the Institute of Communication Networks and Satellite Communications of TU Graz, funded by the Technical University. The outdoor system consists of an azimuth/elevation tracking pedestal, a 3\,m parabolic dish with feed, low-noise amplifier and converter for the S-Band downlink, and a cross-Yagi antenna for the UHF uplink. Located in the nearby control room are: (1) the program-track antenna control system, (2) a 700\,W UHF transmitter, (3) the demodulator for the downlink signal and (4) the BRITE terminal node controller (TNC), handling communications with the spacecraft. Five computer consoles are available to the operators, plus a large screen showing the current location of the spacecraft. Two consoles are dedicated to commanding the spacecraft and displaying telemetry. The other screens are used for monitoring of the signal strength (via a spectrum analyzer), mission activity logging, displaying the spacecraft attitude, and for teleconferences with the other BRITE teams. 

All received data (telemetry and raw science data) are stored on a server which is accessible by the science teams. The station can be operated fully automatically and also via remote access. Typically 6 to 7 passes are tracked per day; half of them in the early morning hours and the other half in the evening. Station operations to date have been very smooth and reliable. The BRITE signal strength is excellent and in most cases, data download is carried out at the maximum rate of 256\,kbit\,s$^{-1}$.

\begin{figure}[h] 
\center
\includegraphics[width=0.6\textwidth]{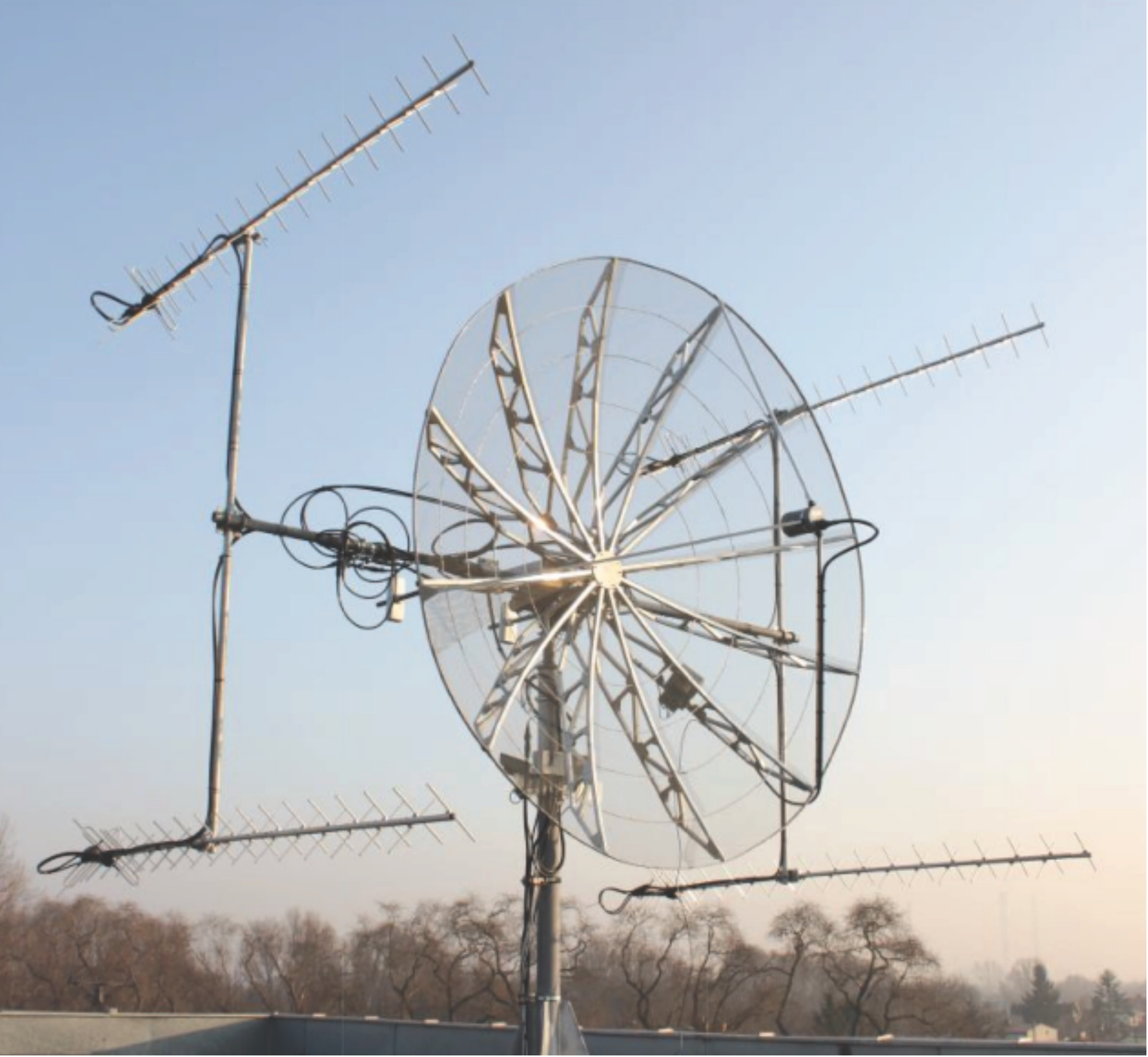}
\caption{Ground station antennas at the Copernicus Astronomical Center, Warsaw} 
\label{fig-TUGround4}
\end{figure}

The Warsaw Ground Station is fully operational as the main station for communication with the Polish BRITE satellites and it may serve also as backup for Austria, and vice versa. On several occasions, the downlink signals from BRITE-Austria and Lem were received simultaneously by the Warsaw and Graz ground stations with large power margins. Only at very low elevation angles do close-by cell phone towers cause shortlived interference in some directions. Filters are being installed in the pre-amplifiers to deal with this situation.

\begin{figure}[h] 
\center
\includegraphics[width=1\textwidth]{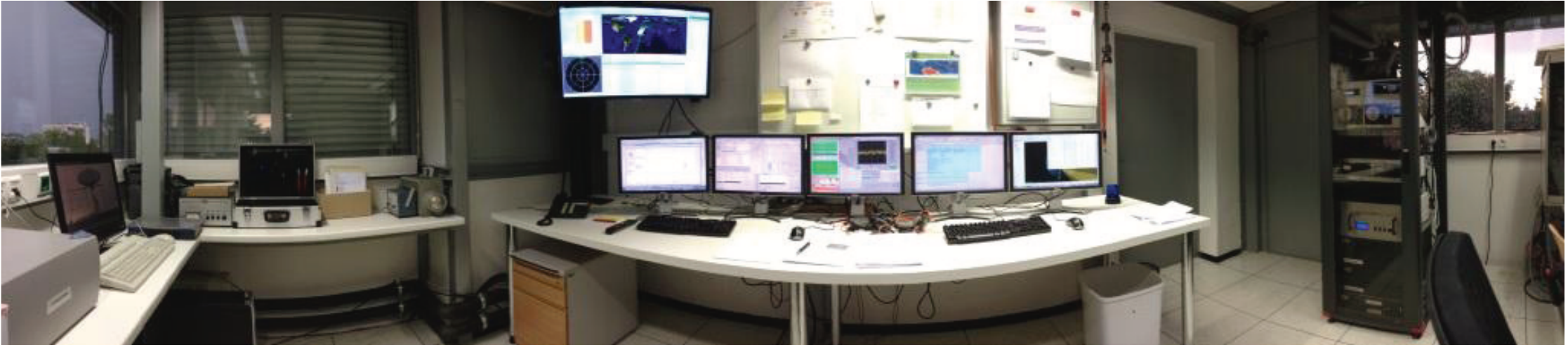}
\caption{Control room for the ground station at  Graz University of Technology} 
\label{fig-TUGround3}
\end{figure}

Fig.\,\ref{fig-TUGround4} shows the antenna system of the ground station at the Copernicus Astronomical Center in Warsaw, and Fig.\,\ref{fig-TUGround3} the control room of a similar ground station located at the Graz University of Technology, Austria. 

Beginning in October 2013, during commissioning of the Austrian BRITE nanosats, an unexpected problem appeared: strong interference from sources located in Northern Europe which reduce the data throughput as fewer commands and acknowledgements on the UHF uplink get through to the satellite. These strong sources of interference are detected on all three BRITE satellites whenever passing over Europe. They seem to extend from northern Europe (Scotland), through central Europe (Mukhachevo) to the  southeast (Cyprus). Detailed signal readings have been made by all three BRITEs and mitigation measures are underway.

The ground station used to communicate with UniBRITE from Toronto (SFL) is the same as that used for MOST and two other already orbiting nanosatellites. The UHF uplink is provided by a roof-mounted set of four cross-yagi antennas while the S-band downlink is provided via a tower mounted 2\,m parabolic dish antenna. 

\subsection{Mission management, target selection and data policy}  \label{ss:management}   

The operations and control of each individual BRITE nanosat are the responsibility of each Principal Investigator (PI) and his institution which receives funding from the respective national agency.

The PIs are part of the BRITE Executive Science Team (BEST), which oversees: (a) evaluation of target proposals, (b) generation of the observing schedules and strategies, (c) data processing, reduction and archiving, (d) allocation of data sets to users, (e) coordinating BRITE publications and conference presentations, and (f) overall management and control of the project.  There are two voting members per participating nanosat from each country. The BEST members are identified in the list of co-authors of this paper with a ``$\dagger$". 

BEST meets twice a year to review mission status, to provide a progress report on the scientific activities arising from \bc\ data, to evaluate input from the community, to review the observing program and to consider any changes to the mission plan, if deemed appropriate or necessary.  The \bc\ Data Archive is located at the Copernicus Astronomical Center (Warsaw, Poland) and is responsible for communicating the science data to the BRITE community.

Although \bc\ is currently funded and operated by three countries, cooperation with the broader international community was encouraged from the beginning, with the first Announcement of Opportunity for BRITE science proposals issued on 30 September 2008. An international BRITE conference was held in Vienna in 2009, and various BRITE workshops have been held since then in Poland, Canada, and Austria. Further information, including policies on data rights and handling, can be found at  http://www.brite-constellation.at and the links therein, and at  http://www.univie.ac.at/tops/CoAst/archive/cia152.pdf the proceedings of the 2009 Vienna conference.  Meanwhile, BEST has received observing proposals which were compiled in the \bc\ Target Star Catalogue. This catalogue now includes every star brighter than V\,=\,4\,mag, with a significant fraction as faint as V\,=\,6\,mag.  Some targets have been proposed independently by up to 5 different scientists. Figure\,\ref{fig-sky} gives a sense of the distribution on the sky of the \bc\ Target Star Catalogue.

\section{Commissioning of the first nanosats}     \label{s:comm}    

The first contact with BRITE-Austria was successfully established and the healthy state of the satellite was confirmed only 3 hours after launch in 2013 during the first pass over the Graz ground station. During the next pass, contact was made with UniBRITE and its health was confirmed. 
Soon after these initial contacts, SFL began on-orbit commissioning of UniBRITE and the Graz University of Technology (with technical support from SFL) began commissioning of BRITE-Austria. 

From the initial radio contacts with the satellite, it was clear that the performance of the new v4 UHF uplink receiver was substantially better than that of the earlier version of the receiver. The S-band downlink transmitter (of which a prototype had flown previously) met its specifications.

After release of the Austrian nanosats from the launch vehicle into orbit, they tumbled as expected. Once the detumbling procedure (using only data from the magnetometer and magnetorquers) was initiated in less than one orbit and sooner than expected, the rotation rates of both spacecraft were reduced to the point where coarse three-axis pointing was possible. This Coarse Mode of pointing uses the sun sensors, magnetometer, star tracker and reaction wheels. Fine Pointing Mode (using the startracker and reaction wheels only) was achieved for UniBRITE in early July 2013. A full-frame image of the field was taken by the instrument and an observation was performed with three $70\times 70$ pixel rasters. Subsequent observations with good pointing used the starfield maps generated by the BRITE-Target software. 

After the three onboard computers were checked out, the 6 sun sensors (which aid in coarse attitude determination by sensing the direction of the Sun) and the magnetometers were successfully commissioned without anomalies. On-orbit magnetometer recalibration in late June 2013 reduced an error related to the local magnetic field vector from $20\deg$ to less than $5\deg$. This was sufficient to allow convenient transition to satellite fine pointing. 

The three reaction wheels, whose pointing performance is crucial if BRITE is to produce highly precise photometry, were also tested.  The startracker, which is used to most accurately determine the satellite's orientation by observing a pre-selected star field, required fine-tuning of its various setup parameters and some modifications of the original operating software. The startrackers were having more difficulty than planned to find stable quaternions. Diagnosing the reasons for this, and finding a solution, extended the commissioning time beyond the nominal schedule. To maximize efficiency, UniBRITE commissioning was focused on optimizing fine pointing, while commissioning of BRITE-Austria was dedicated to characterizing the instrument performance. The first image of the first commissioning star field was taken on 23 March 2013 by BRITE-Austria.

About five weeks after launch, a problem with the ADCS computer was detected on BRITE-Austria, likely due to a cosmic ray hit when the nanosat passed through the South Atlantic Anomaly (SAA), a ``dimple" in the Earth's magnetosphere where satellites in low Earth orbit are exposed to higher-than-average particle fluxes. The external memory could no longer be correctly read out. It was decided to re-arrange the software such that the housekeeping computer runs both the housekeeping and the ADCS software. This is a redundancy in the BRITE design  where both computers have access to all sensors and actuators. The new software was provided by SFL, tested in the laboratory, and successfully uploaded to the satellite from the Graz ground station. Stable pointing was achieved shortly afterwards. The new software running on the housekeeping computer turned out to simplify operations and this hybrid version has been installed on all BRITE nanosats.

Another problem area emerged in the cosmetics of the CCDs on both nanosats.  In the first test exposures, both CCDs exhibited an unusually high number of ``warm" pixels and some column defects.  Repeated exposures over several weeks revealed that the warm pixel incidence and time evolution is consistent with damage due to radiation exposure.  To explore the problem, to develop strategies (such as ``annealing" the CCDs and adopting optimized timing and readout protocols), and to propose changes in the designs of the Canadian and Polish BRITE nanosats while they were still on the ground, BEST created a BRITE CCD Tiger Team, chaired by Stephan Mochnacki. The column defects are removable by subtracting the median of every column from every pixel in that column. This correction may be done on the spacecraft before data are selected and transmitted. Hot pixels -- which at present represent about 2\% of all pixels -- are mapped and corrected also via median filtering with neighboring good pixels.  A BRITE Photometry Tiger Team (PHOTT) was recently created, chaired first by Rainer Kuschnig and presently by Bert Pablo and Gemma Whittaker, to coordinate efforts to optimize  the extraction of photometric signals from \bc\ data.

Based on the spacecraft commissioning and early science commissioning of the instruments, a summary of the performances of UniBRITE and BRITE-Austria is: \\
- Pointing accuracy and stability exceeds the mission requirements: $\sigma\,=\,1.5$ pixel (i.e. 40 arcsec) over 15 days, based on 6030 subframes and frequent re-pointings (see Fig.\,\ref{fig-pointing}). \\
- Photometric noise in $\tilde{R}$ and $\tilde{B}$ for stars with V\,=\,4\,mag presently is 1.5\,mmag for a 15-min long series of exposures, but the PHOTT Tiger Team predicts that a precision of 1\,mmag in 15 minutes is achievable with an improved instrument  model, meeting the mission requirement. This noise level is likely to be improved when we get longer observing intervals per satellite orbit and higher cadences. 
For the least symmetric PSFs the $\sl rms$ scatter is  larger. \\
- A linear regression for 14 stars ranging in spectral type from O9 to A3 results in  \\ 
\hspace*{10mm}V\,=\,1.059($\pm 0.044)\,\cdot\,\tilde{R}$ + offset. \\ 
A better determination of this and of a colour term must await measurements of a larger sample of stars covering a broader range of spectral types. \\
- Radiation degradation of the CCDs on orbit is obvious, but changes to respond to this sensitivity in the three BRITE nanosats yet to be launched are underway. Tests with radiation sources and cyclotron particle beams were performed with engineering-grade CCDs and various shielding materials, as well as tests of the devices at different voltage levels and temperatures. For the CCDs already in orbit, on average, the degradation rate is just under 0.02\% of the pixels per year, or 1.5 pixels per month per subraster, on a level ranging from 4$\sigma$ above noise level to saturation. PHOTT has put forward strategies to largely correct the photometry for the CCD cosmetic effects. \\
- The shortest exposure time has a data cadence of 0.1 second, with an overhead of 6.5 seconds for processing (independent of the exposure time). \\
- High-precision flatfield factors can be obtained from empty fields as the satellites drift between target fields during each orbit. \\
- Intra-pixel variations have been measured in all pixels and we are working on correction strategies. \\
- Stray light is not a concern so far. We obtained good photometry even when the Moon was close to the field. \\
- We have experience only with fields containing several bright stars, hence, about $\pm 20^{\rm o}$ from the Galactic Equator. The planned procedure of visiting a series of fields sequentially during an orbit will be tested after all primary commissioning goals have been completed.

\section{Conclusions}

\bc\ opens a new era for nanosatellites, which initially were seen primarily as student projects for education. We have demonstrated that nanosatellites have potential for high-level astrophysical research, provided that innovative technology is used; in the case of BRITE, high-precision 3-axis stabilisation of a very-low-mass payload. Nano- and microsatellites should be designed, built and tested according to the principle that reliability increases and cost decreases when design complexity decreases.  The reduced cost means that individual universities and small research institutions can compete with large agencies in well selected research niches.

\bc\ is a highly successful partnership between three countries (Austria, Canada and Poland) which is open to other international partners. One scientifically rewarding (but technically challenging) opportunity of an expanded partnership would be an expanded Constellation with an expanded wavelength range, extending into the ultraviolet and infrared.

\acknowledgements{
The authors wish to thank the Canadian Space Agency (CSA) for its financial support of two concept studies and funding of two Canadian BRITE nanosats, to the Ontario Centres of Excellence Inc. (ETech Division) for their support of the development of enabling nano-satellite technologies, and to the Natural Sciences \& Engineering Research Council (NSERC) Canada for general financial support. They also thank the University of Vienna, the Technical University of Graz,  the Austrian Research Promotion Agency (FFG/ALR), and the Austrian Ministry of Transport, Innovation and Technology (bmvit) for financing and leading the development of the first two BRITE satellites. Polish participation in the BRITE project was made possible by a generous grant (427/FNiTP/78/2010) by the Polish Ministry of Science and Higher Education to the participating institutes of the Polish Academy of Science.  The authors express their thanks to Jagoda Daszy\'nska for her contributions to plans for BRITE stellar seismology, and to the Technical University of Vienna for contributing a remotely controlled ground station. Funding assistance is gratefully acknowledged by the Canadian authors to the Canadian Space Agency in support of HQP: Flights for the Advancement of Science and Technology (FAST), and by the Austrian authors to the Austria Science Fonds (FFG P22691-N16). KZ acknowledges support from the Fund for Scientific Research of Flanders (FWO), Belgium, under grant agreement G.0B69.13, and by the Polish authors to the NCN grant 2011/01/M/ST9/05914.
}


\end{document}